\documentstyle[preprint,aps]{revtex}

\tightenlines

\newcommand{\brho}{\mbox{\boldmath $\rho$}}

\newcommand{\be}{\begin{equation}}
\newcommand{\ee}{\end{equation}}

\newcommand{\bea}{\begin{eqnarray}}
\newcommand{\eea}{\end{eqnarray}}

\newcommand{\bx}{\mbox{${\bf x}$}}
\newcommand{\by}{\mbox{${\bf y}$}}
\newcommand{\bz}{\mbox{${\bf z}$}}
\newcommand{\bxpr}{\mbox{${\bf x}^{\prime}$}}
\newcommand{\bypr}{\mbox{${\bf y}^{\prime}$}}
\newcommand{\ct}{\mbox{${\cal T}$}}
\newcommand{\co}{\mbox{${\cal O}$}}
\newcommand{\cD}{\mbox{${\cal D}$}}
\newcommand{\bR}{\mbox{${\bf R}$}}
\newcommand{\bB}{\mbox{${\bf B}$}}
\newcommand{\bE}{\mbox{${\bf E}$}}
\newcommand{\bA}{\mbox{${\bf A}$}}
\newcommand{\bF}{\mbox{${\bf F}$}}

\newcommand{\Rb}{\mbox{$\overline{R}$}}
\newcommand{\wc}{\mbox{$\omega_{c}$}}
\newcommand{\wb}{\mbox{$\omega_{b}$}}

\begin{document}
\draft

\title{Berry's Phase in the Presence of a Stochastically Evolving
        Environment: A Geometric Mechanism for Energy-Level Broadening}

\author{Frank Gaitan}
\address{Department of Physics; Boston College; Chestnut Hill, MA 02167-3811}
\date{\today}

\maketitle

\begin{abstract}
The generic Berry phase scenario in which a two-level system is coupled to 
a second system whose dynamical coordinate is slowly-varying is generalized to 
allow for stochastic evolution of the slow system. The stochastic behavior is
produced by coupling the slow system to a heat reservoir which is 
modeled by a bath of harmonic oscillators initially in equilibrium at
temperature $T$, and whose spectral density has a bandwidth which is 
small compared to the energy-level spacing of the fast system. The 
well-known energy-level shifts produced by Berry's phase in the fast system, 
in conjunction with the stochastic motion of the slow system, leads to a 
broadening of the fast
system energy-levels. In the limit of strong damping and sufficiently low 
temperature, we determine the degree of level-broadening analytically, and
show that the slow system dynamics satisfies a Langevin equation in which
Lorentz-like and electric-like forces appear as a consequence of geometrical
effects. We also determine the average energy level-shift produced in the fast 
system by this mechanism.
\end{abstract}


\section{Introduction}
\label{sec1}
In the 14 years since its discovery, Berry's phase has proven to be a
fruitful development in our understanding of the adiabatic limit of quantum
mechanics \cite{sha}. In his original analysis \cite{ber}, Berry considered 
a quantum system with a discrete, non-degenerate energy spectrum whose
dynamics is driven by a set of classical parameters ${\bf R}(t)$ that
vary slowly on the time scale of the quantum system. He showed that when 
${\bf R}(t)$
was cycled adiabatically through a loop in parameter space, and the quantum
system was initially prepared in an eigenstate $|E(0)\rangle$ of the
initial Hamiltonian $H(0)$, the quantum system returned to the initial
state $|E(0)\rangle$ at the end of the cycle, to within a phase factor
$\exp(i\phi)$. This much was in agreement with the quantum adiabatic theorem.
What was new was that the phase $\phi$ contained a contribution $\phi_{g}$
whose origin was found to be deeply geometrical \cite{sim}, and
which had been discarded in previous treatments of this theorem. $\phi_{g}$ is
now referred to as Berry's phase, and is a functional $\phi_{g}[{\bf R}(t)]$
of the loop traced out by ${\bf R}(t)$ in parameter space.

Not long after Berry's discovery, it was suggested by Moody et.\ al.\ 
\cite{mod} that Berry's phase should be observable in nuclear magnetic
resonance. It was argued that Berry's phase would produce a shift in the
energy levels of the resonating spin which would alter the observable
resonant frequencies. This shift was subsequently observed by Suter et.\ al.\
\cite{sut}. Berry's phase has also been observed in a number of other
experimental settings \cite{sha}. For our purposes, though, the NMR context
will be the most relevant. 

The original Berry scenario was generalized in a number of ways in the
years following his discovery \cite{sha}. For the purposes of this paper,
however, the most interesting development was the elevation of the slowly
varying classical parameters ${\bf R}(t)$ to the status of dynamical
quantum variables \cite{br2,me2}. The total system now consists of a slow
subsystem with coordinates ${\bf R}(t)$ coupled to a fast spin-like degree
of freedom $\mbox{\boldmath $\sigma$}$. The separation of dynamic time
scales allows a Born-Oppenheimer treatment of the coupled dynamics. The
fast system is again found to develop a Berry phase in those states evolving
out of an instantaneous energy eigenstate due to the motion of the slow system. 
Berry's phase again produces shifts in the fast system energy levels. 
These shifts are functionals $\delta E
[{\bf R}(t)]$ of the slow system path ${\bf R}(t)$. Geometrical effects also
influence the motion of the slow system. They lead to the appearance 
of gauge fields which act on the motion of the slow 
system \cite{br2,me2,jac,mea,zyg}.
The gauge fields produce Lorentz-like and electric-like 
forces in the classical equation of motion of the slow system.

In this paper, we examine a generalization of the Born-Oppenheimer scenario
in which the slow system is allowed to evolve stochastically. The stochastic
behavior is introduced by coupling the slow system to a heat resevoir
which we represent by a bath of harmonic oscillators initially in thermal
equilibrium at temperature $T$. At first, all 3 subsystems will be quantum
mechanical. Ultimately, we will be interested in the case where the stochastic 
behavior is classical, so eventually we will take the semiclassical limit of 
the coupled slow system/heat resevoir dynamics. The fast system remains quantum
mechanical throughout. We will examine how the stochastic motion of
the slow system influences the Berry phase effects discussed above.

For easy reference, we summarize our principal results. {\em First}, 
because the semiclassical motion of the slow system is a random process,
the energy-level shifts produced by Berry's phase in the fast system's 
spectrum will cause these levels to broaden. This can be simply understood 
in the following way. Due
to the stochastic force acting on the slow system (see below), the
boundary conditions imposed on the path ${\bf R}(t)$ no longer determine 
a unique path. Instead, all possible realizations of the random process
${\bf R}_{i}(t)$ must be considered which satisfy the boundary conditions.
Each realization ${\bf R}_{i}(t)$ will produce a level-shift
$\delta E[{\bf R}_{i}(t)]$ with a probability $P[{\bf R}_{i}(t)]$.
Thus all possible shifts in the original energy $E$ are produced, though
with differing probabilities. The result is that the original energy level
$E$ is broadened out by the stochastic evolution of the slow system,
in conjunction with the Berry phase induced level-shifts. We will 
calculate the average energy shift $\overline{\delta E}$ produced, and 
the variance associated with the spread of energy shifts about 
$\overline{\delta E}$,
under appropriate restrictions. {\em Second}, we show that the semiclassical
motion of the slow system is governed by a Langevin equation. Geometrical
effects are seen to produce the same Lorentz-like and electric-like forces in 
this equation as were found in the original Born-Oppenheimer scenario
in which the slow system evolved deterministically. We are able to determine 
the probability distribution $P[{\bf R}_{i}(t)]$ which we use to calculate the
average energy level-shift and broadening produced in the fast system.

The organization of this paper is as follows. We begin in Section~\ref{sec2} 
by reviewing those Berry phase results which are pertinent to the
work described in this paper. In Section~\ref{sec3} we introduce the model 
system we shall study. In Section~\ref{sec4} we set-up a path integral 
representation of its dynamics. We implement the Born-Oppenheimer 
approximation, watching carefully for
the appearance of Berry's phase. We also integrate out the unobserved
reservoir degrees of freedom. This yields an effective description of the
slow system dynamics in which only ${\bf R}(t)$ enters as a dynamical 
variable. Although the fast system and the resevoir no longer appear in this 
effective description explicitly, they do produce a back action on the 
effective slow dynamics which we obtain while carrying out the above maneuvers. 
In Section~\ref{sec5}
we take the semiclassical limit of the effective slow system dynamics. We
find that the slow system motion is governed by a Langevin equation in which
the geometry-induced gauge forces appear. We also obtain the probability
distribution governing the random process ${\bf R}(t)$. In Section~\ref{sec6}
we use this probability distribution to set-up a generating function for
the moments of the energy-level shift $\delta E[{\bf R}(t)]$. We use
the generating function to calculate the average energy-level shift and 
level-broadening, under appropriate restrictions. In Section~\ref{sec7}, we 
summarize our results and make closing remarks. Finally, in an Appendix, we
calculate the spectral density for classical Brownian motion; the goal
being to determine what restrictions are needed to insure that
the stochastic motion of the slow system is adiabatic relative to the fast 
system.

\section{Berry Phase Preliminaries}
\label{sec2}
In this section we collect the Berry phase results which are relevant to the
work we shall present below. The reader is referred to Ref.~\cite{br2}
for a more detailed presentation.

In the original Berry phase scenario \cite{ber}, the focus of attention is
a quantum system with a discrete, non-degenerate energy spectrum. Its
Hamiltonian $H[{\bf R}]$ is assumed to depend on a set of classical
parameters ${\bf R}$ which represent an environmental degree of
freedom to which the quantum system is coupled. The environment is assumed
to evolve adiabatically. This produces an adiabatic time dependence in the
quantum Hamiltonian, $H=H[{\bf R}(t)]$. The time dependence of the 
quantum state $|\psi (t) \rangle$ is determined by solving Schrodinger's 
equation using the quantum adiabatic theorem. Towards this end, one introduces 
the energy eigenstates of the instantaneous Hamiltonian $H[{\bf R}(t)]$,
\begin{displaymath}
H[{\bf R}(t)]\: |E[{\bf R}(t)]\: \rangle = E[{\bf R}(t)]\:|E[{\bf R}(t)]
   \: \rangle \hspace{0.4in} .
\end{displaymath}
It is further assumed that the environment is taken adiabatically around a 
loop in parameter space such that ${\bf R}(T) ={\bf R}(0)$, and that the 
quantum system is initially prepared in an
eigenstate $|E[{\bf R}(0)]\rangle$ of the initial Hamiltonian $H[{\bf R}(0)]$.
The quantum adiabatic theorem states that, at time $t$, the quantum system 
will be found in the state $|E[{\bf R}(t)]\rangle$ to within a phase factor,
\begin{equation}
|\psi(t)\rangle = \exp\left[ i\gamma_{E}(t) -\frac{i}{\hbar}
    \int_{0}^{t}\, d\tau E[{\bf R}(\tau)]\right] |E[{\bf R}(t)]\rangle
      \hspace{0.1in} . \label{psit}
\end{equation}
The second term in the phase of the exponential is known as the dynamical phase
and was already familiar from previous studies of the quantum adiabatic 
theorem. The first term represents Berry's discovery, and is referred to as 
Berry's phase, 
\begin{equation}
\gamma_{E}(t) = i \int_{0}^{t}\, d\tau\,\langle E[{\bf R}(\tau)]|
                 \frac{\partial}{\partial\tau}|E[{\bf R}(\tau)]\rangle
                  \hspace{0.4in} . \label{bert}
\end{equation}
In the cases where Berry's phase is physically relevant, $\gamma_{E}$ is
non-integrable: it cannot be written as a single-valued function of ${\bf R}$
over all of parameter space. Simon \cite{sim} showed that the quantum adiabatic
theorem has a line bundle structure inherent in it, and that Schrodinger's
equation defines a parallel transport of the quantum state 
around the line bundle. Berry's phase is the signature that the associated 
connection has non-vanishing curvature.

Berry's phase has been observed in a number of physical systems \cite{sha}.
For present purposes, the nuclear magnetic resonance experiments are
the most interesting. Moody et.\ al.\ \cite{mod} pointed out that Berry's
phase should alter the observed resonant frequencies in NMR. In particular,
if one were examining the resonance associated with the pair of levels
$E$ and $E^{\prime}$, the shift in the resonant frequency $\Delta\omega_{0}$ 
would be,
\begin{displaymath}
\Delta\omega_{0} = \frac{\gamma_{E}(\ct)-\gamma_{E^{\prime}}(\ct)}{\ct}
    \hspace{0.4in} ,
\end{displaymath}
where $\ct$ is the period of the oscillating transverse magnetic field
${\bf H}_{\perp}(t)$. The resonant frequency shift occurs because a shift
$\delta E$ is produced in each energy-level $E$ by Berry's phase,
\begin{equation}
\delta E = \frac{\hbar\gamma_{E}(\ct)}{\ct} \hspace{0.4in} . \label{shift}
\end{equation}
In the NMR experiments both $E$ and $\dot{\gamma}_{E}$ were time independent.
Since $\gamma_{E}(\ct)$ is independent of the parameterization of ${\bf R}(t)$,
so long as the new parameterization remains adiabatic, eqn.~(\ref{shift})
will also be correct when $\dot{\gamma}_{E} \neq constant$. One simply 
reparameterizes the time $t \rightarrow t^{\prime}$ in such a way that 
$\dot{\gamma}_{E}(t^{\prime})$ is time independent.

The above scenario can be generalized. We promote ${\bf R}(t)$ from a
classical degree of freedom with no dynamics of its own, to a fully-dynamical
quantum variable \cite{br2,me2}. As above, ${\bf R}(t)$ couples to a
spin-like degree of freedom $\mbox{\boldmath $\sigma$}$. For the
remainder of this paper we will assume that $\mbox{\boldmath $\sigma$}$ 
corresponds to a pseudo-spin 1/2. To stay within the context of the quantum 
adiabatic theorem,
we assume that a Born-Oppenheimer treatment of the coupled dynamics is
appropriate, and that ${\bf R}$ ($\mbox{\boldmath $\sigma$}$) is the dynamical
variable of the slow (fast) system. The Hamiltonian for the coupled system
is taken to be
\begin{equation}
H_{tot} = \frac{{\bf P}^{2}}{2M} + V[{\bf R}]
                  - g{\bf R}\cdot\mbox{\boldmath $\sigma$} 
                   \hspace{0.4in} .
\end{equation}
Here ${\bf P}$ is the momentum conjugate to ${\bf R}$; 
$\mbox{\boldmath $\sigma$}_{i}$ are the Pauli matrices; 
we allow for the possibility of a potential $V$ acting on the slow system; 
and $g$ is a coupling constant.

In applying the Born-Oppenheimer approximation, one first considers the fast
system, treating the slowly-varying ${\bf R}$ as fixed. The fast motion is 
governed by $H_{f}= -g{\bf R}\cdot\mbox{\boldmath $\sigma$}$ with eigenstates 
$|E_{\pm}({\bf R})\rangle$ and energies $\pm gR$, where $R=|{\bf R}|$. In 
fact, ${\bf R}$ varies adiabatically. From the quantum adiabatic theorem we 
know that when the fast system is prepared initially in the state $|E[{\bf R}
(0)]\rangle$, its state at time $t$ is given by eqns.~(\ref{psit}) and 
(\ref{bert}), and Berry's phase produces the energy level shift
$\delta E$ given by eqn.~(\ref{shift}). The slow dynamics is governed
by
\begin{eqnarray}
H_{eff} & = & \langle E({\bf R})|H_{tot}|E({\bf R})\rangle \nonumber \\
   & = & \frac{\left({\bf P}-{\bf A}[{\bf R}]\right)^{2}}{2M}
    +\Phi\left({\bf R}\right) + E\left({\bf R}\right)
      \hspace{0.1in} .
\label{heff}
\end{eqnarray}
Here,
\begin{equation}
{\bf A}({\bf R}) = i\hbar\langle E({\bf R})|\nabla_{{\bf R}}|E({\bf R})\rangle
  \hspace{0.4in} , \label{ar}
\end{equation}
and,
\begin{equation}
\Phi({\bf R}) = \frac{\hbar^{2}}{2}\sum_{i=\pm} g_{ii}({\bf R}) 
    \hspace{0.4in} .
\label{phir}
\end {equation}
From eqns.~(\ref{bert}) and (\ref{ar}), we see that ${\bf A}[{\bf R}]$ is 
related to Berry's phase: 
$\hbar\dot{\gamma}_{E}= \dot{{\bf R}}\cdot{\bf A}[{\bf R}]$. $g_{ij}({\bf R})$
is the quantum metric tensor \cite{br2,pro} and corresponds to the real
part of the quantum geometric tensor $T_{ij}$,
\begin{displaymath}
T_{ij}=\langle \partial_{i} E|\left( 1- |E\rangle\langle E|\right)
        |\partial_{j} E\rangle \hspace{0.4in} ,
\end{displaymath}
where $\partial_{i}=\partial/\partial R_{i}$. We see that geometrical
effects produce gauge potentials $\Phi$ and ${\bf A}$ in the effective
Hamiltonian that governs the slow dynamics. In the semiclassical limit, the
equation of motion for the slow system is \cite{br2}
\begin{equation}
M\ddot{{\bf R}}= \dot{{\bf R}}\times{\bf B}[{\bf R}] -\nabla_{{\bf R}}
                  \left[\Phi[{\bf R}] +  V[{\bf R}] +
                   E[{\bf R}]\right] ,
\label{brnopp}
\end{equation}
where ${\bf B} = \nabla\times{\bf A}$. Thus, geometrical effects lead
to the appearance of Lorentz-like and electric-like forces that act on the
slow system in the semiclassical limit.

This concludes our summary of the Berry phase physics we shall need below.

\section{The Model}
\label{sec3}
As mentioned in the Introduction, we would like to explore how the Berry
phase physics discussed in Section~\ref{sec2} is affected by stochastic
motion of the slow system. To produce stochastic behavior, 
we introduce a heat resevoir that couples to the slow system, and
which initially is in thermal equilibrium at temperature $T$. Because
the resevoir degrees of freedom are unobserved, they must be traced out
of the dynamical description. The result of this operation is that stochastic
and frictional forces appear in the dynamics of the slow system.

To produce a tractable model, some restrictions must be placed on the heat
resevoir, and on its coupling to the slow system. These restrictions, and
the model to which they lead, have been discussed in great detail in the
literature \cite{cl1,fey}. Consequently, our discussion here will be brief.
The reader is referred to these papers for further discussion.

The first restriction imposed is that the heat resevoir contain an infinite 
number of degrees of freedom. The idea here is that any energy transferred 
from the slow system to the infinitely many resevoir degrees of freedom 
effectively disappears into the resevoir, not to return 
in any physically reasonable amount of time. The result is dissipation of 
energy and irreversibility in the motion of the slow system.

The second restriction is that the slow system couple weakly to the
individual resevoir degrees of freedom. As such, each resevoir degree
of freedom is only weakly disturbed from equilibrium. This restriction
allows us to analyze the excitation of the resevoir away from equilibrium
using a harmonic approximation \cite{cl1,fey}. Consequently, the
resevoir degrees of freedom can be modeled by harmonic oscillators that
represent the normal modes of the resevoir.

This combination of restrictions is known to produce stochastic
and frictional forces in the semiclassical motion of the slow system
once the resevoir has been traced out \cite{cl2,sch}. Because Berry's
phase requires the slow system to evolve adiabatically relative to the
fast system, we must add a further restriction to insure adiabaticity
of the stochastic motion. The essential point is the following. Because
of the stochastic force, ${\bf R}(t)$ contains fluctuations or noise.
The range of frequencies $\Delta\omega$ present in the fluctuations
can be determined by examining the spectral density $J_{{\bf R}}(\omega)$ of
${\bf R}(t)$. In the Appendix we show that, in the semiclassical limit,
$J_{{\bf R}}(\omega)$ is proportional to the spectral density 
$J_{{\bf F}}(\omega)$ of the stochastic force ${\bf F}(t)$. The range of
frequencies (bandwidth) present in ${\bf F}(t)$ is of order $1/\tau_{c}$,
where $\tau_{c}$ is the correlation time of ${\bf F}(t)$.
The correlation time is known for our model of the resevoir \cite{sch}:
$\tau_{c}=\hbar /\sqrt{6}kT$. Thus the bandwidth $\Delta\omega$ of the 
random process ${\bf R}(t)$ is also of order $1/\tau_{c}$. We can insure 
that ${\bf R}(t)$ evolves adiabatically if we require that $\hbar\Delta\omega$ 
be much less than the energy-level spacing of the fast system, $\Delta E =
E_{+}-E_{-}$. This is the final restriction we need to impose on the
slow system/reservoir dynamics.
For example, since $\hbar\Delta\omega\sim\sqrt{6}kT$, we see that 
when $1\: {\rm K} \leq T \leq 100\: {\rm K}$, the bandwidth
for fluctuations in ${\bf R}(t)$ satisfies
$10^{-4}\: {\rm eV} \leq \hbar\Delta\omega \leq 10^{-2}\: {\rm eV}$.
Thus if $\Delta E > 0.1 \: 
{\rm eV}$, the slow system will evolve adiabatically relative to the fast
system, even though its dynamics is stochastic.

Assuming that all 3 of these restrictions are satisfied, the following
model Hamiltonian will produce adiabatic stochastic motion of the slow system: 
\begin{eqnarray}
H & = & \left(\frac{{\bf P}^{2}}{2M} + V[{\bf R}]\right) 
   - g{\bf R}\cdot
         \mbox{\boldmath $\sigma$} \nonumber \\
  &   &  {} \hspace{0.1in} +\sum_{j}\frac{m_{j}}{2}\left( \dot{{\bf Q}}_{j}^{2}
          -\omega^{2}_{j}{\bf Q}_{j}^{2}\right) \nonumber \\
  &   & {} \hspace{0.3in} + \sum_{j} C_{j}{\bf R}
             \cdot {\bf Q}_{j} 
             + {\bf R}^{2}\sum_{j}\frac{C_{j}^{2}}{2m_{j}
             \omega^{2}_{j}} \hspace{0.1in} .
\label{hcomp}
\end{eqnarray}
The first term on the right-hand side (rhs) is the non-interacting
slow system Hamiltonian $H_{s}$. The second term on the rhs is the
slow system/fast system interaction Hamiltonian $H_{sf}$. The third term
is the non-interacting resevoir Hamiltonian $H_{r}$, and the last two terms
give the slow system/resevoir interaction Hamiltonian $H_{sr}$. 
$m_{j}$ and $\omega_{j}$ are the oscillator masses and
frequencies, respectively; and $C_{j}$ are the coupling constants for
the slow system/oscillator interactions. These constants are not determined
by the model, though they satisfy a constraint \cite{cl1} that connects them 
to the friction coefficient $\eta$ that will appear in 
the Langevin equation obtained in Section~\ref{sec5}. The last term on the 
rhs is a counter-term introduced to insure that $V({\bf R})$ is the potential
acting on the slow system in the semiclassical limit.

Having introduced our model, our next task is to set-up a path integral
treatment of its dynamics.
 
\section{Dynamics via Path Integrals}
\label{sec4}
In this Section we apply the influence functional formalism of
Feynman and Vernon \cite{fey} to describe the dynamics of our model.
The analysis begins with the density matrix for the composite system
(slow system/fast system/resevoir). The essentials of its time development
are presented in Section~\ref{sec4a}. The Born-Oppenheimer approximation
is implemented in Section~\ref{sec4b}, and the resevoir degrees of freedom
are traced out in Section~\ref{sec4c}. The main result of this section
will be a path integral representation of the reduced density matrix
describing the effective dynamics of the slow system. Geometrical 
contributions to this effective dynamics appear during implementation
of the Born-Oppenheimer approximation; stochastic and dissipative effects
enter in when tracing out the resevoir, though this will not become apparent 
until Section~\ref{sec5}.

\subsection{Density Matrix Preliminaries}
\label{sec4a}
We begin with the density operator $\rho(t)$ of the composite system. Its
evolution is given by,
\begin{equation}
\rho(t)   = 
  \exp\left[ -\frac{i}{\hbar}\left(t + \frac{{\cal T}}{2}\right)
           H\right] \rho\left(-\frac{{\cal T}}{2}\right)  
 \times\exp\left[
           \frac{i}{\hbar}\left(t+\frac{{\cal T}}{2}\right) H \right] .
\label{dop}
\end{equation}
$H$ is the Hamiltonian for the composite system (see eqn.~(\ref{hcomp})), and 
the time evolution begins at $t=-{\cal T}/2$.

It proves useful to work in the basis $| {\bf x},\:\sigma ,\: 
{\bf z}\,\rangle$ in which the slow system is at ${\bf x}$, the fast system
has pseudo-spin projection $\sigma$ along ${\bf x}$, and the harmonic
oscillators have positions ${\bf Z}=\, (\ldots , {\bf Z}_{j}, \ldots )$,
where $j$ labels the oscillator degrees of freedom. In this basis,
eqn.~(\ref{dop}) becomes,
\begin{eqnarray}
\lefteqn{\langle\, {\bf x}_{f}, \sigma_{f}, {\bf Q}_{f}|\,\rho(t)\, | 
    {\bf y}_{f},
  \sigma_{f}^{\prime}, {\bf Z}_{f}\,\rangle = } \nonumber \\
& & {} \hspace{0.6in}   \sum_{\sigma ,\sigma^{\prime}} 
   \int\, d{\bf x}^{\prime}\,
   d{\bf y}^{\prime}\, d{\bf Q}^{\prime}\, d{\bf Z}^{\prime}
    \:\langle {\bf x}^{\prime},\sigma ,{\bf Q}^{\prime}
     |\rho (-{\cal T}/2)|{\bf y}^{\prime}, \sigma^{\prime},
      {\bf Z}^{\prime}\rangle \nonumber \\
& & {} \hspace{1.20in} \times K({\bf x}_{f},\sigma_{f},{\bf Q}_{f}, t; 
        {\bf x}^{\prime},\sigma ,
     {\bf Q}^{\prime}, -{\cal T}/2) 
 \: K^{\ast}( {\bf y}_{f},\sigma_{f}^{\prime},
         {\bf Z}_{f},t;
     {\bf y}^{\prime},\sigma^{\prime},{\bf Z}^{\prime},-{\cal T}2) .
\label{dmat}
\end{eqnarray}
Here,
\begin{displaymath}
K( {\bf x}_{f}, \sigma_{f}, {\bf Q}_{f},t; {\bf x}^{\prime},\sigma ,
  {\bf Q}^{\prime}, -{\cal T}/2) =  
 \langle {\bf x}_{f}, \sigma_{f},{\bf Q}_{f} | \exp\left[
    -\frac{i}{\hbar}\left( t + \frac{{\cal T}}{2}\right) H \right]
     | {\bf x}^{\prime},\sigma , {\bf Q}^{\prime} \rangle .
\end{displaymath}
$K^{\ast}$ is the complex conjugate of $K$ and will not be written out
explicitly. In Section~\ref{sec4b} we will obtain a path integral expression
for $K$.

We must specify an initial condition for the time development. We assume
that the resevoir is uncoupled from the slow/fast systems prior to
$t=-{\cal T}/2$. Thus $\rho(-{\cal T}/2)$ factors:
\begin{equation}
\rho\left( -{\cal T}/2\right) = \rho_{r}\left( -{\cal T}/2
  \right)\,\rho_{sf}\left( - {\cal T}/2\right) \hspace{0.1in} .
\end{equation}
The resevoir is assumed to be in thermal equilibrium at temperature $T$
initially.
Thus the energy-levels of each oscillator degree of freedom are populated
according to the Boltzmann distribution, and $\rho_{r}(-{\cal T}/2)$ is
the product of the density matrices of the individual oscillators \cite{fey}.
The initial condition on $\rho_{sf}(-{\cal T}/2)$ is that it correspond
to a Born-Oppenheimer pure state. Thus the slow system is prepared in the
position eigenstate $| {\bf x}^{\prime} \rangle$, and the fast system in the
negative energy state $|E_{-}[{\bf x}^{\prime}]\rangle$ corresponding to
the pseudo-spin aligned along ${\bf x}^{\prime}$ (i.~e.\ $\sigma = 1/2$): 
$\rho_{sf}(-{\cal T}/2)=|{\bf x}^{\prime}, 1/2 \rangle\langle {\bf x}^{\prime}
, 1/2| \equiv \tilde{\rho}_{E_{-}}(-{\cal T}/2)$. Thus,
\begin{eqnarray}
\lefteqn{\langle {\bf x}_{f},\sigma_{f}, {\bf Q}_{f}|\rho(t)|{\bf y}_{f},
  \sigma_{f}^{\prime}, {\bf Z}_{f}\rangle = } \nonumber \\
&  & {} \hspace{0.6in}\int\, d{\bf x}^{\prime}
   d{\bf y}^{\prime} d{\bf Q}^{\prime} d{\bf Z}^{\prime} 
    \:\rho_{r}( {\bf Q}^{\prime}, {\bf Z}^{\prime},
      -{\cal T}/2) \:\tilde{\rho}_{E_{-}}( {\bf x}^{\prime},
       {\bf y}^{\prime}, - {\cal T}/2) \nonumber \\
&  & {} \hspace{1.0in} \times K( {\bf x}_{f}, \sigma_{f}, {\bf Q}_{f}, t; 
    {\bf x}^{\prime},
     1/2, {\bf Q}^{\prime}, -{\cal T}/2)
\: K^{\ast}( {\bf y}_{f}, \sigma_{f}^{\prime}, 
      {\bf z}_{f}, t;
     {\bf y}^{\prime}, 1/2, {\bf Z}^{\prime},-{\cal T}/2) . 
\label{fprop}
\end{eqnarray}

\subsection{Born-Oppenheimer Treatment of Slow/Fast Dynamics}
\label{sec4b}
Implementing the Born-Oppenheimer approximation in Berry phase systems is
well-understood \cite{br3,kur,gai}. We present the analysis for $K$. A
similar analysis applies for $K^{\ast}$, but will not be given here.

$K$ describes evolution over the time interval [$-{\cal T}/2,\: t$]. We
introduce intermediate times $t_{i}$, ($i= 1, \ldots , N-1$), and break
the evolution into $N$ smaller intervals of duration $\epsilon = ( t +
{\cal T}/2)/N$. Thus,
\begin{equation}
K = \langle {\bf x}_{f},\sigma_{f},{\bf Q}_{f}|\prod_{i=1}^{N} U(t_{i},
     t_{i-1}) | {\bf x}^{\prime}, 1/2, {\bf Q}^{\prime}\rangle ,
\label{kprop}
\end{equation}
with $U(t_{i},t_{i-1}) = 1 -i\epsilon H(t_{i})/\hbar $. Inserting basis
sets $|{\bf x}_{i},\sigma_{i},{\bf Q}_{i}^{\prime}\rangle \equiv  |{\bf x}_{i}
\rangle\, |E_{\sigma_{i}}[{\bf x}_{i}]\,\rangle |{\bf Q}_{i}^{\prime}\rangle$ 
at the intermediate times $t_{i}$, eqn.~(\ref{kprop}) breaks up into a product 
of factors describing evolution over the time intervals ($t_{i-1}$, $t_{i}$).
The factor associated with the $i$-th time interval is,
\begin{eqnarray}
\lefteqn{\langle {\bf x}_{i},\sigma_{i},{\bf Q}^{\prime}_{i}|
\left[ 1- \frac{i\epsilon}{\hbar} 
 (H_{s}+H_{sf}+H_{r}+H_{sr})\right]
  |{\bf x}_{i-1},\sigma_{i-1},{\bf Q}^{\prime}_{i-1}\rangle } 
    \nonumber \\
 & {} \hspace{1.25in} = & 
       {} \hspace{0.2in} \langle {\bf x}_{i}, {\bf Q}^{\prime}_{i}|
     \left[ \begin{array}{l}
               \langle\, E_{\sigma_{i}}[{\bf x}_{i}]| \left( 1 
                - \frac{i\epsilon}{\hbar} H_{s}\right) |E_{\sigma_{i-1}}
                  [{\bf x}_{i-1}]\,\rangle \\
               {} \hspace{0.3in} -\frac{i\epsilon}{\hbar}E_{\sigma_{i}}
                [{\bf x}_{i}]\,\delta_{\sigma_{i},\sigma_{i-1}} \\
     {} \hspace{0.5in} -\frac{i\epsilon}{\hbar} (H_{r}+H_{sr}) 
          \delta_{\sigma_{i},\sigma_{i-1}} + {\cal O}(\epsilon^{2})
            \end{array} \right]
      |{\bf x}_{i-1}, {\bf Q}^{\prime}_{i-1}\rangle
\end{eqnarray}
Recall that the slow system evolves adiabatically relative to the fast
system. Thus the fast system will remain in the initial (negative)
energy-level (i.\ e.\ $\sigma_{i}=1/2$ for all $i$). Furthermore
\cite{br2,br3,kur,gai},
\begin{displaymath}
\langle\, E_{\sigma_{i}}[{\bf x}_{i}]|E_{\sigma_{i-1}}[{\bf x}_{i-1}]\,
  \rangle = 1 + i\epsilon\dot{\gamma}_{-}(t_{i}) + {\cal O}(\epsilon^{2}),
\end{displaymath}
and,
\begin{displaymath}
\langle\, E_{\sigma_{i}}[{\bf x}_{i}]|\, H_{s}\,|E_{\sigma_{i-1}}
  [{\bf x}_{i-1}]\, \rangle = H_{eff}(t_{i}) + {\cal O}(\epsilon^{2}) .
\end{displaymath}
Here $\gamma_{-}(t)$ is the Berry phase associated with the fast system state
$|E_{-}[{\bf x}]\,\rangle$, and $H_{eff}(t)$ is given by 
eqn.~(\ref{heff}), with the state appearing in eqns.~(\ref{ar}) and 
(\ref{phir}) given by $|E_{-}[{\bf x}]\,\rangle$. 

Inserting all these results into eqn.~(\ref{kprop}) gives, 
\begin{equation}
K = \langle {\bf x}_{f}, {\bf Q}_{f}|\exp\left[ -\frac{i}{\hbar}
   \int_{-{\cal T}/2}^{t} d\tau H^{\prime}(\tau ) \right]
    |{\bf x}^{\prime}, {\bf Q}^{\prime}\rangle ,
\label{nkprop}
\end{equation}
where,
\begin{displaymath}
H^{\prime}(\tau) = -\hbar\dot{\gamma}_{-}(\tau ) + E_{-}(\tau) +
                    H_{eff}(\tau) .
\end{displaymath}
Notice that explicit reference to the fast system has disappeared because
the adiabatic time dependence and the initial condition force it to remain
in the $E_{-}$ energy-level at all times. In its place, the back action
of the fast system on the slow system has produced the Born-Oppenheimer
potential energy $E_{-}[{\bf x}]$, and the gauge potentials $\Phi [{\bf x}]$
and ${\bf A}[{\bf x}]$ in $H^{\prime}(\tau)$. Equation~(\ref{nkprop}) can be 
written as a path integral in the usual way \cite{slm}, 
\begin{equation}
K({\bf x}_{f},\sigma_{f},{\bf Q}_{f},t;{\bf x}^{\prime}, 1/2, 
  {\bf Q}^{\prime} , -{\cal T}/2) = 
\delta_{\sigma_{f},1/2}\int {\cal D}{\bf x}(t)
   {\cal D}{\bf Q}(t)\exp\left[\frac{i}{\hbar}(S_{s}+S_{r}+S_{sr})\right] .
\label{path}
\end{equation}
Here,
\begin{eqnarray}
S_{s} & = & \int dt \left[ \frac{M}{2}\dot{{\bf x}}^{2} + \dot{{\bf x}}\cdot
         {\bf A}[{\bf x}] - \Phi [{\bf x}]-E_{-}[{\bf x}]-V[{\bf x}] \right]
\label{sslow}            ,  \\
S_{r} & = & {}\hspace{0.0in} \int dt \sum_{j}\frac{m_{j}}{2}\left( 
           \dot{{\bf Q}}^{2}_{j} -
          \omega_{j}^{2}{\bf Q}_{j}^{2}\right),  \\
S_{sr}& = & {} \hspace{0.0in}  \int dt\left[ -\sum_{j}C_{j}{\bf x}_{j}\cdot 
            {\bf Q}_{j}
          -{\bf x}^{2}\sum_{j}\frac{C_{j}^{2}}{2m_{j}\omega_{j}^{2}}
           \right] .  
\end{eqnarray}
All paths ${\bf x}(t)$ (${\bf Q}(t)$) appearing in eqn.~(\ref{path}) begin at
${\bf x}^{\prime}$ (${\bf Q}^{\prime}$) and end at ${\bf x}_{f}$
(${\bf Q}_{f}$).

\subsection{Tracing Out the Resevoir}
\label{sec4c}
When $K$ and $K^{\ast}$ are substituted into eqn.~(\ref{fprop}), we find, 
\begin{eqnarray}
\lefteqn{\langle {\bf x}_{f},\sigma_{f}, {\bf Q}_{f}|\rho(t)|{\bf y}_{f},
  \sigma^{\prime}_{f}, {\bf Z}_{f} \rangle = 
 \delta_{\sigma_{f},1/2}\,\delta_{\sigma_{f}^{\prime},1/2} }
\nonumber \\
& & {} \hspace{1.2in} \times\int d{\bf x}^{\prime}d{\bf y}^{\prime}
        \tilde{\rho}_{E_{-}}({\bf x}^{\prime},{\bf y}^{\prime}, -{\cal T}/2) 
    \nonumber \\
& & {} \hspace{1.5in} \times\int d{\bf Q}^{\prime}d{\bf Z}^{\prime}  
          \rho_{r}({\bf Q}^{\prime},{\bf Z}^{\prime}, -{\cal T}/2) 
   \nonumber \\
& & {} \hspace{1.8in} \times\int {\cal D}{\bf x}(t)
       {\cal D}{\bf x}^{\prime}(t) {\cal D}{\bf Q}(t) 
        {\cal D}{\bf Q}^{\prime}(t) 
 \exp\left[\frac{i}{\hbar}\left( S[{\bf x}, {\bf Q}]
         -S[{\bf x}^{\prime},{\bf Q}^{\prime}]\right)\right] .
\label{nnprop}
\end{eqnarray}
Since the resevoir degrees of freedom are unobserved, we must trace over
them. Setting ${\bf Q}_{f}={\bf Z}_{f}$ in eqn.~(\ref{nnprop}) and
integrating over ${\bf Q}_{f}$, the left-hand side (lhs) becomes the reduced 
density matrix $\tilde{\rho}_{E_{-}}({\bf x}_{f},{\bf y}_{f}, t)$,
while the full equation determines its time dependence,
\begin{equation}
\tilde{\rho}_{E_{-}}({\bf x}_{f},{\bf y}_{f},t) 
  = \int d{\bf x}^{\prime}d{\bf y}^{\prime} J_{E_{-}}({\bf x}_{f},
        {\bf y}_{f},t; {\bf x}^{\prime}, {\bf y}^{\prime}, -{\cal T}/2)
         \,\tilde{\rho}_{E_{-}}({\bf x}^{\prime}, {\bf y}^{\prime},-
          {\cal T}/2),
\label{bigr}
\end{equation}
where,
\begin{equation}
J_{E_{-}}({\bf x}_{f},{\bf y}_{f},t;{\bf x}^{\prime},{\bf y}^{\prime},
           -{\cal T}/2) 
 = \int {\cal D}{\bf x}(t) {\cal D}{\bf x}^{\prime}(t) \exp\left[
       \frac{i}{\hbar}\left(S_{s}[{\bf x}]-S_{s}[{\bf x}^{\prime}]\right)
        \right]{\cal F}[{\bf x}, {\bf x}^{\prime}] .
\label{temp}
\end{equation}
All paths ${\bf x}(t)$ (${\bf x}^{\prime}(t)$) appearing in eqn.~(\ref{temp})
begin at ${\bf x}^{\prime}$ (${\bf y}^{\prime}$) and end at ${\bf x}_{f}$
(${\bf y}_{f}$).
${\cal F}[{\bf x},{\bf x}^{\prime}]$ is the influence functional \cite{fey}
which contains all the effects of the resevoir on the motion of the slow
system. For a resevoir composed of harmonic oscillators, ${\cal F}[{\bf x},
{\bf x}^{\prime}]$ can be evaluated exactly \cite{fey},
\begin{equation}
{\cal F}[{\bf x}, {\bf x}^{\prime}] 
 = \exp\left[ -\frac{1}{\hbar}\int_{-{\cal T}/2}^{t} d\tau ds \left[
      {\bf x}(\tau)-{\bf x}^{\prime}(\tau)\right]\,\left[ \alpha(\tau - s)
       {\bf x}(s) - \alpha^{\ast}(\tau - s){\bf x}^{\prime}(s)\right]\right],
\end{equation}
and $\alpha(\tau - s) = \alpha_{R}(\tau - s)+i\alpha_{I}(\tau -s)$ (see below).
Thus,
\begin{eqnarray}
\lefteqn{J_{E_{-}}({\bf x}_{f}, {\bf y}_{f},t;{\bf x}^{\prime},
   {\bf y}^{\prime}, -{\cal T}/2) = 
    \int {\cal D}{\bf x}(t) {\cal D}{\bf x}^{\prime}(t) } \nonumber \\
& & {} \hspace{0.6in} \times 
      \exp\left[\frac{i}{\hbar}\left( S_{s}[{\bf x}]-
          S_{s}[{\bf x}^{\prime}] - \int_{-{\cal T}/2}^{t} d\tau ds
           \left[ {\bf x}(\tau)-{\bf x}^{\prime}(\tau)\right]\alpha_{I}
            (\tau - s)\left[ {\bf x}(s)+{\bf x}^{\prime}(s)\right]\right)
             \right] \nonumber \\
& & {} \hspace{1.7in} \times\exp\left[-\frac{1}{\hbar}
        \int_{{\cal T}/2}^{t} d\tau ds \left[ {\bf x}(\tau)-{\bf x}^{\prime}
         (\tau)\right] \alpha_{R}(\tau - s)\left[ {\bf x}(s) -
          {\bf x}^{\prime}(s)\right]\right] ,
\label{Jprop}
\end{eqnarray}
where,
\begin{eqnarray}
\alpha_{R}(\tau -s) & = & \sum_{j}\frac{C_{j}^{2}}{2m_{j}\omega_{j}}
                         \coth\left(\frac{\hbar\omega_{j}}{2kT}\right)
                          \cos\left[\omega_{j}(\tau -s)\right] \\
\alpha_{I}(\tau -s) & = & -\sum_{j}\frac{C_{j}^{2}}{2m_{j}\omega_{j}}
                           \sin\left[\omega_{j}(\tau -s)\right] .
\label{aimag}
\end{eqnarray}
Eqns.~(\ref{Jprop})--(\ref{aimag}) give the effective dynamics of the slow
system. Geometrical effects have produced gauge fields in $S_{s}[{\bf x}]$;
and the stochastic and dissipative effects due to the resevoir lurk in the
terms containing $\alpha_{R}$ and $\alpha_{I}$, respectively, though this
will not become apparent until we take the semiclassical limit of the 
effective dynamics.

\section{Slow System Effective Dynamics: Semiclassical Limit}
\label{sec5}
Within the context of our model, we have established the slow system's
effective quantum dynamics. In this Section we will see that the 
semi-classical limit of this dynamics is dissipative and stochastic.
This limit is taken in Section~\ref{sec5a}. In Section~\ref{sec5b} we
obtain the Langevin equation governing the stochastic motion, and the
probability distribution functional characterizing the statistical
properties of this motion. The Langevin equation is found to contain 
the same geometric forces that appeared in the deterministic
Born-Oppenheimer scenario discussed in Section~\ref{sec2}.

\subsection{Semiclassical Limit of $J_{E_{-}}$}
\label{sec5a}
Eqn.~(\ref{temp}) describes the slow system effective dynamics. In the
absence of a resevoir, the influence functional 
${\cal F}[\bx ,\bxpr ] =1$, and the
semiclassical limit of the remaining exponential factor can be obtained by
the method of steepest descent. When the resevoir is present, 
${\cal F}[\bx ,\bxpr ] \neq 1$, and we must examine the form it takes 
when $\hbar \rightarrow 0$.
For a resevoir composed of harmonic oscillators, this limit has been
carried out in Ref.~\cite{cl2}. The result is, 
\begin{eqnarray}
\lefteqn{J^{sc}_{E_{-}}({\bf x}_{f},{\bf y}_{f},t;{\bf x}^{\prime},
  {\bf y}^{\prime},-{\cal T}/2) = \int {\cal D}{\bf x}(t){\cal D}
{\bf x}^{\prime}(t) } \nonumber \\
& & {} \hspace{0.9in} \times\exp\left[\,\frac{i}{\hbar}\left(\, S_{s}[{\bf x}]
        -S_{s}[{\bf x}^{\prime}] -\frac{\eta}{2}\int_{-{\cal T}/2}^{t}\,
          d\tau\,\left[ {\bf x}(\tau)-{\bf x}^{\prime}(\tau)\right]\cdot
    \left[ \dot{{\bf x}}(\tau)+\dot{{\bf x}}^{\prime}(\tau)\right]\right)
            \right] \nonumber \\
& & {} \hspace{1.6in} \times\exp\left[\, -\frac{\eta KT}{\hbar^{2}}
        \int_{-{\cal T}/2}^{t}\, d\tau\,\left\{ {\bf x}(\tau) - 
         {\bf x}^{\prime}(\tau)\right\}^{2}\,\right] \hspace{0.1in} .
\label{dyno}
\end{eqnarray}
The physical significance of $\eta$ will become clear in Section~\ref{sec5b}.

The real exponential factor in eqn.~(\ref{dyno}) introduces important 
simplifications when $\hbar \rightarrow 0$. In this limit, it is very
sharply peaked about ${\bf x}(\tau) = {\bf x}^{\prime}(\tau)$ so that
the dominant contribution to the integral comes from pairs of paths for
which $|{\bf x}(\tau) - {\bf x}^{\prime}(\tau)| \ll 1$, for all $\tau$.
Because the paths $\bx (\tau)$ and $\bxpr (\tau)$ are continuous,
$J_{E_{-}}^{sc}$ is also sharply peaked about ${\bf x}_{f}={\bf y}_{f}$
and ${\bf x}^{\prime}={\bf y}^{\prime}$. By eqn.~(\ref{bigr}),
$\rho_{E_{-}}^{sc}({\bf x},\: {\bf y},\: t)$ is similarly sharply peaked about 
${\bf x}={\bf y}$. Thus, expanding the argument of
the complex exponential in eqn.~(\ref{dyno}) to second order in
[${\bf x}(\tau)-{\bf x}^{\prime}(\tau)$] introduces negligible error
when $\hbar\rightarrow 0$ \cite{sch}. To carry out the expansion, it proves
convenient to introduce center-of-mass and relative coordinates, respectively,
${\bf R}(\tau)$ and $\mbox{\boldmath $\rho$}(\tau)$: 
\begin{equation}
{\bf R}(\tau) = \frac{{\bf x}(\tau)+{\bf x}^{\prime}(\tau)}{2}
   \hspace{0.5in} ; \hspace{0.5in} \brho(\tau) = {\bf x}(\tau) - 
                                    {\bf x}^{\prime}(\tau) \hspace{0.1in} .
\label{comrel}
\end{equation}
Using eqn.~(\ref{sslow}), it is straightforward to show that
\be
S_{s}[\bx ]-S_{s}[\bxpr ]= \int_{-{\cal T} /2}^{t}\, d\tau\,\left[\brho (\tau)
 \cdot \left( -M\ddot{\bR}+\dot{\bR}\times\bB[\bR ]+\bE[\bR ]\right) +
  \co\left( |\brho |^{3}\right) \right] .
\ee
Here $\bB [\bR ] = \nabla\times\bA [\bR ]$; $\bE [\bR ] = -\nabla\left(
\Phi [\bR ]+ V[\bR ] + E_{-}[\bR ]\right)$; and $\bA [\bR ]$ and
$\Phi [\bR ]$ are the geometric gauge potentials discussed in 
Section~\ref{sec2}. It follows immediately from eqn.~(\ref{comrel}) that,
\be
-\frac{\eta}{2}\int_{-{\cal T} /2}^{t}\, d\tau\,\left[ \bx (\tau)-\bxpr (\tau)
  \right]\cdot\left[\dot{\bx}(\tau)+\dot{\bx}^{\prime}(\tau)\right] =
   -\eta \int_{-{\cal T} /2}^{t}\, d\tau\, \brho (\tau)\cdot\dot{\bR}(\tau) .
\ee
Putting together all these results gives,
\bea
\lefteqn{J_{E_{-}}^{sc}(\bx_{f},\by_{f},t;\bxpr ,\bypr ,-\ct /2) = 
  \int \cD\bR (t) \cD\brho (t) } \nonumber \\
& & {} \hspace{2.0in} \times\exp\left[ \frac{i}{\hbar}\int_{-{\cal T} /2}^{t}
         d\tau \brho (\tau)\cdot\left[ -M\ddot{\bR}-\eta\dot{\bR}+
          \dot{\bR}\times\bB + \bE \right]\right] \nonumber \\
& & {} \hspace{2.8in} \times\exp\left[ -\frac{\eta kT}{\hbar^{2}}
          \int_{-{\cal T} /2}^{t} d\tau\left[\brho (\tau)\right]^{2}\right] .
\label{stoch}
\eea

\subsection{Appearance of Stochastic Dynamics}
\label{sec5b}
To bring out the stochastic character of the dynamics contained in
eqn.~(\ref{stoch}), we again focus on the real exponential in this
equation. Its argument can be re-written as,
\be
-\frac{\eta kT}{\hbar^{2}}\int_{-{\cal T}/2}^{t} d\tau \brho (\tau)\cdot\brho
  (\tau) = 
  -\frac{1}{2\hbar^{2}}\int_{-{\cal T}/2}^{t}\int_{-{\cal T}/2}^{t}
   d\tau ds \,\rho_{i}(\tau) A_{ij}(\tau -s)\rho_{j}(s) 
 ,
\ee
where $A_{ij}(\tau -s) = 2\eta kT \delta_{ij}\delta (\tau -s)$. To bring
out the meaning of this term we introduce a Hubbard-Stratonovitch
transformation:
\bea
\Phi[\brho ] & \equiv &  \exp\left[
  -\frac{1}{2\hbar^{2}}\int_{-{\cal T}/2}^{t}\int_{-{\cal T}/2}^{t}
   d\tau ds \rho_{i}(\tau) A_{ij}(\tau -s)\rho_{j}(s) 
    \right] \nonumber \\
& = & \frac{1}{N} \int \cD\bF (t) \exp\left[
  -\frac{1}{2}\int_{-{\cal T}/2}^{t}\int_{-{\cal T}/2}^{t}
   d\tau ds F_{i}(\tau) A_{ij}^{-1}(\tau -s)F_{j}(s) 
    \right] \nonumber \\
& & {} \hspace{1.2in} \times\exp\left[\frac{i}{\hbar}\int_{-{\cal T}/2}^{t} 
        d\tau \brho (\tau) \cdot \bF (\tau) \right] ,
\label{charf}
\eea
where $A_{ij}^{-1}(\tau -s) = (1/2\eta kT)\delta_{ij}\delta (\tau -s)$, and
$N$ is an (infinite) normalization constant which we suppress below. 
Eqn.~(\ref{charf}) is an identity which can be proved by evaluating
the Gaussian integral on the rhs. From eqn.~(\ref{charf}), it is clear
that $\Phi [\brho ]$ is the characteristic functional for the Gaussian
random process $\bF (t)$. Functional derivatives of $\Phi [\brho ]$
with respect to $\brho (t)$ generate the correlation functions of $\bF (t)$.
In particular,
\be
\langle F_{i}(\tau) F_{j}(s)\rangle = \left(\frac{\hbar}{i}\right)^{2}
   \frac{\delta^{2} \Phi [\brho ]}{\delta\rho_{i}(\tau)\delta\rho_{j}(s)}
  = A_{ij}(\tau -s) = 2\eta kT \delta_{ij}\delta (\tau -s) .
\ee
Thus $\bF (t)$ has a two-point correlation function identical to that of a
classical stochastic force produced by a heat resevoir with friction
coefficient $\eta$ and temperature $T$. We shall see below that this is the
correct interpretation for $\bF (t)$.

Making use of eqn.~(\ref{charf}) in eqn.~(\ref{stoch}) gives,
\bea
\lefteqn{J_{E_{-}}^{sc}(\bx_{f},\by_{f},t; \bxpr , \bypr , -\ct /2) = 
\int \cD\bR (t)\cD\brho (t)\cD\bF (t) } \nonumber \\
& & {} \hspace{1.6in} \times\exp\left[ -\frac{1}{4\eta kT}
     \int_{-{\cal T}/2}^{t} d\tau \left[ \bF (\tau) \right]^{2} \right]
      \exp\left[\frac{i}{\hbar}\int_{-{\cal T}/2}^{t} d\tau \brho (\tau)
       \cdot \hat{L}\bR (\tau)\right] ,
\label{stoch2}
\eea
where $\hat{L}\bR (\tau) \equiv -M\ddot{\bR} -\eta\dot{\bR}+\dot{\bR}
\times\bB + \bE$. We recognize the $\brho (t)$-integral as the Dirac
delta functional. Thus,
\bea
\lefteqn{J_{E_{-}}^{sc}(\bx_{f},\by_{f},t;\bxpr ,\bypr ,-\ct /2) =  
 \int\cD\bR (t)\cD\bF (t) } \nonumber \\
& & {} \hspace{2.4in} \times\exp\left[ -\frac{1}{4\eta kT}
      \int_{-{\cal T}/2}^{t}d\tau\left[ \bF (\tau)\right]^{2}\right]\,
       \delta\left[ \hat{L}\bR (t) +\bF (t) \right] .
\eea
Clearly, in the semiclassical limit, the only paths that contribute to
$J_{E_{-}}^{sc}$ are those which satisfy the Langevin equation,
\be
\hat{L}\bR (t) + \bF (t) = 0 ,
\ee
or,
\be
M\ddot{\bR} = -\eta\dot{\bR}+\dot{\bR}\times\bB [\bR ] +\bE [\bR ] +\bF (t) .
\label{lang}
\ee
As promised in Section~\ref{sec3}, introducing the resevoir has caused
the semiclassical dynamics of the slow system to become stochastic. $\eta$
can be interpreted as a friction coefficient, and $\bF (t)$ as a Gaussian
stochastic force. We also see that the geometric Lorentz-like and
electric-like forces which appear in the deterministic Born-Oppenheimer
scenario (see eqn.~(\ref{brnopp})) 
also appear in the (adiabatic) stochastic generalization of this
scenario. Eqn.~(\ref{lang}) corresponds to one of the two main results of
this Section.

To obtain the second, we carry out the $\bF (t)$-integration,
\be
J_{E_{-}}^{sc}(\bR_{f},\brho_{f}=0,t;\bR^{\prime},\brho^{\prime}=0,-\ct /2)
 = \int\cD\bR (t) \exp\left[ -\frac{1}{4\eta kT}\int_{-{\cal T}/2}^{t} d\tau
    \left[ \hat{L}\bR (\tau) \right]^{2}\right] .
\label{stoch3}
\ee
Inserting this into eqn.~(\ref{bigr}), we find,
\bea
\lefteqn{\rho_{E_{-}}^{sc}(\bR_{f}, \brho_{f}=0,t)= } \nonumber \\
& & {} \hspace{0.7in} \int d\bR^{\prime} \int\cD\bR (t) 
   \exp\left[ -\frac{1}{4\eta kT}
   \int_{-{\cal T}/2}^{t} d\tau \left[\hat{L}\bR (\tau)\right]^{2}\right]
    \rho_{E_{-}}^{sc}(\bR^{\prime},\brho^{\prime}=0,-\ct /2) .
\label{stoch4}
\eea
We see that $\rho_{E_{-}}^{sc}(\bR_{f},t)$ is found by summing over all paths
which lead from $\bR^{\prime}$ to $\bR_{f}$, and then integrating over all 
possible $\bR^{\prime}$. From eqn.~(\ref{stoch4}), the probability that a 
path $\bR (t)$ lies in
a ``volume'' $\cD\bR (t)$ in the space of paths that join $\bR^{\prime}$
to $\bR_{f}$ is clearly,
\be
\cD P = \exp\left[ -\frac{1}{4\eta kT}\int_{-{\cal T}/2}^{t} d\tau\left[
         \hat{L}\bR (\tau)\right]^{2}\right] \cD\bR (t) .
\ee
This identifies the probability distribution functional $P[\bR (t)]$ as,
\be
\label{probdis}
P[\bR (t)] = \exp\left[ -\frac{1}{4\eta kT}\int_{-{\cal T}/2}^{t} d\tau
              \left[ \hat{L}\bR (\tau)\right]^{2}\right] .
\ee
We will use eqn.~(\ref{probdis}) in the following section to calculate the
average shift and broadening of the fast system energy-levels.

\section{Fast System Energy Level-Shift and Broadening} 
\label{sec6}
In this Section we calculate approximately the average shift and broadening 
produced 
in the fast system energy levels by the combination of Berry phase and
stochastic effects. In Section~\ref{sec6a}, we introduce the generating 
function that is the basis of our calculation, and state the conditions
we must impose to make the calculation tractable. The generating function
is evaluated in Section~\ref{sec6b}. In Section~\ref{sec6c} and \ref{sec6d},
we determine the approximate average level-shift and level-broadening, 
respectively.

\subsection{Preliminaries}
\label{sec6a}
We restrict the slow system to two-spatial dimensions: $\bR (t)= R(t)
[\cos\phi (t),\:\sin\phi (t), 0]$. $H_{sf}$ takes the form:
\begin{displaymath}
H_{sf} = gR(t)\left( \begin{array}{lr}
                        0         & e^{-i\phi} \\
                        e^{i\phi} & 0
                     \end{array} \right) .
\end{displaymath}
We assume throughout this Section that the fast system is initially prepared
in the state $| E_{-}[\bR_{i}]\,\rangle$, where $\bR (-\ct /2)\equiv \bR_{i}$.
Using eqns.~(\ref{ar}) and (\ref{phir}), 
\be
\Phi_{-}[\bR ] = \frac{(\hbar\nabla\phi)^{2}}{2M}=\frac{\hbar^{2}}{2MR^{2}} 
   \hspace{0.7in} ; \hspace{0.7in} 
 \bA [\bR ] =-\frac{\hbar\nabla\phi}{2} = \frac{\hbar}{2R^{2}}\bR
    \times\hat{\bz} .
\label{bdefs}
\ee
From these, $\bB =\nabla\times\bA$, and $\bE = -\nabla(\Phi_{-}[\bR] +
V[\bR ]+E_{-}[\bR ])$. To simplify the following calculation, we choose
$V[\bR ] = -E_{-}[\bR ]$.

Our Langevin equation becomes,
\be
M\ddot{\bR} = -\eta\dot{\bR} -\frac{h}{2}\delta^{2}(\bR ) \dot{\bR}\times
  \hat{\bz} + \frac{h^{2}}{MR^{3}}\hat{\bR} + \bF (t) . 
\label{2dlang}
\ee
The Lorentz force is seen to vanish everywhere except at the origin, while
the electric force is divergent there and repulsive. Consequently, any
finite energy motion of the slow system must have a turning point with
$R_{tp}\neq 0$. In 2D, then, the Lorentz force disappears from 
eqn.~(\ref{2dlang}). The electric force is seen to cut off rather quickly,
and is second order in Planck's constant $h$. Thus, away from the origin, the 
electric force is small in the semiclassical limit. To simplify the following
analysis, we replace the exact electric force with something qualitatively
similar. To capture the strong repulsion at the origin, we simply remove
the origin from the xy-plane so that $R(t)\neq 0$ for all $t$. To mimic
the rapid cutoff away from the origin, we simply drop the electric force
from eqn.~(\ref{2dlang}). With this simplification, our Langevin equation
reduces to free Brownian motion in the punctured xy-plane:
\be
\hat{L}\bR = M\ddot{\bR} + \eta\dot{\bR} = \bF .
\label{free}
\ee
The boundary conditions are: $\bR (-\ct /2)=\bR_{i}$, and $\bR (\ct /2)=
\bR_{f}$.

We write $\bR (t)=\bR_{0}(t) + \by (t)$. Here $\bR_{0}(t)$ is the homogeneous
solution of eqn.~(\ref{free}); and $\by (t)$ is a particular solution.
$\bR_{0}(t)$ satisfies the same boundary conditions as
$\bR (t)$, so consequently, $\by (\pm \ct /2) =0$. We want a condition
that insures that the fluctuating particular solution $\by (t)$ remains
small compared to the noiseless homogeneous solution $\bR_{0}(t)$ in an 
average sense: $\langle |\by (t)|^{2}\rangle \ll |\bR_{0}(t)|^{2}$. For
Brownian motion, it is well-known that \cite{ein}, 
\be
\langle |\by (t)|^{2}\rangle \sim \frac{2kT}{\eta} t .
\label{albert}
\ee
Defining $\overline{{\bf R}}=(\bR_{i}+\bR_{f})/2$; requiring 
$\langle |\by (t)|^{2}\rangle \ll \Rb^{2}$; and using eqn.~(\ref{albert})
gives the condition we seek,
\be
kT\ll\frac{\eta\Rb^{2}}{\ct} .
\label{lown}
\ee
Brownian motion which satisfies eqn.~(\ref{lown}) will be referred to as 
low-noise Brownian motion.

The central object of this Section is the generating function $F(\rho )$ for 
the moments of the energy-level shifts
 $\delta E_{-}[\bR ]$ (see eqn.~(\ref{shift})):
\be
F(\rho) = \int \cD\bR (t)\, P[\bR]\exp[\rho\delta E[\bR]] .
\label{genfunc}
\ee
It follows immediately that,
\be
\left. \frac{d[\ln F(\rho) ]}{d\rho}\right|_{\rho = 0}=
  \frac{\int\cD\bR (t)\, P[\bR ]\delta E[\bR ]}{\int\cD\bR (t)\, P[\bR]}
   = \overline{\delta E} ,
\label{ebar}
\ee
and,
\be
\frac{d^{2} [\ln F(\rho )]}{d\rho^{2}}\left. \right|_{\rho = 0} =
  \overline{(\delta E)^{2}}-(\overline{\delta E})^{2} = \sigma^{2} .
\label{var}
\ee
We will use the standard deviation $\sigma$ as a measure of the energy-level 
broadening.

\subsection{Generating Function}
\label{sec6b}
Using eqns.~(\ref{bert}), (\ref{shift}), (\ref{ar}), (\ref{probdis}), 
(\ref{bdefs}), and (\ref{genfunc})
gives, 
\be
F(\rho) = \int\cD\bR (t)\,\exp\left[ -\frac{1}{4\eta kT}
  \int_{-{\cal T}/2}^{{\cal T}/2}\, d\tau\,\left\{ (\hat{L}\bR )^{2} +
    e\rho\,\hat{\bz}\cdot\dot{\bR}\times\bR /R^{2}\right\}\right] ,
\label{genf2}
\ee
where $e= -2\eta kT\hbar /\ct$. It is not possible to evaluate 
eqn.~(\ref{genf2}) analytically under arbitrary conditions. The calculation
is analytically tractable if we restrict ourselves to low-noise Brownian
motion. As discussed in Section~\ref{sec6a}, in the low-noise limit, the
fluctuating component $\by (t)$ is always small compared to the noiseless
component $\bR_{0}(t)$.This allows us to expand the argument $A$ of the 
exponential in eqn.~(\ref{genf2}) 
to second-order in $\by (t)$, yielding a Gaussian path integral. The
presence of $R^{2}$ in the denominator of the second term 
appearing in the integrand of $A$ complicates
the analysis however. Since the noise component $\by (t)$ is small,
$R^{2}\approx \Rb^{2}$ throughout the motion. Consequently, we will
approximate $R^{2}$ by $\Rb^{2}$ in this denominator.
Strictly speaking, we are discarding terms that should be kept by doing this.
As such, our results in Section~\ref{sec6c} and \ref{sec6d} should be 
thought of 
more as rough estimates, than as rigorous results (see the discussion at
the end of Section~\ref{sec6d}). 
In principle, one could repeat the calculation below including these extra 
terms, though we will not do so here. Carrying out this expansion gives,
\be
F(\rho)=\exp\left[\rho\delta E_{-}[\bR_{0}]\right]\int\cD\by (t)\,\exp
         \left[ -\frac{1}{4\eta kT}\int_{-{\cal T}/2}^{{\cal T}/2}
           \, d\tau\, I[\by ]\right] .
\label{genf3}
\ee
Here,
\be
I[\by ]=\left[ M^{2}(\ddot{\by})^{2}+\eta^{2}(\dot{\by})^{2}+2M\eta\,\dot{\by}
         \cdot\ddot{\by}\right] + \frac{e\rho\epsilon^{3ij}}{\Rb^{2}}
          \left[2\dot{R}_{0}^{i}\,y_{j} + \dot{y}_{i}\,y_{j}\right] ,
\label{iy}
\ee
and $\epsilon^{ijk}$ is the Levi-Civita density.

It proves useful to Fourier transform $\by (t)$. A sine-transform is needed
since $\by (\pm\ct /2) = 0$,
\begin{displaymath}
\by (t)=\sum_{n=1}^{n_{b}} \by_{n}\sin\left(\frac{n\pi}{{\cal T}}t\right) .
\end{displaymath}
The need for adiabatic noise requires an upper cutoff 
$\omega_{b}= n_{b}\pi /\ct$ on the noise spectrum
(see Section~\ref{sec3}). The boundary conditions produce a low frequency
cutoff $\omega_{c}=\pi /\ct$. We assume $\ct$ is large, though finite, so
that $\by (t)$  can be written as a complex Fourier integral,
\be
\by (t) = \frac{1}{2\pi}\int_{-\omega_{b}}^{\omega_{b}} d\omega \,\by (\omega)
           \exp[-i\omega t] ,
\label{transf}
\ee
where continuation to negative $\omega$ requires 
$\by (-\omega) =-\by (\omega)$. Note that this is consistent with the 
boundary condition requirement that there be no 
zero-frequency mode in the noise spectrum.
Substituting eqn.~(\ref{transf}) into eqn.~(\ref{genf3}), and completing
the square in the integrand gives (eventually),
\bea
F(\rho) & = & \exp\left[ \rho\delta E_{-}[\bR_{0}] + \frac{1}{4\eta kT}
               \int_{-\wb}^{\wb}\frac{d\omega}{2\pi} B^{\ast}_{i}(\omega)
                A^{-1}_{ij}(\omega)B_{j}(\omega) \right] \nonumber \\
        &   & {} \hspace{0.75in} \times\int\cD\bz (\omega)\exp\left[ -
                     \frac{1}{4\eta kT}\int_{-\wb}^{\wb}\frac{d\omega}{2\pi}
                      z^{\ast}_{i}(\omega)A_{ij}(\omega)z_{j}(\omega)
                       \right] , \label{genf4}
\eea
where,
\bea
A_{ij}(\omega) & = & \delta_{ij}\left( M^{2}\omega^{4} + \eta^{2}\omega^{2}
\label{aij}           \right) + \epsilon^{3ij}\left(
                       \frac{i\omega e\rho}{\Rb^{2}}\right) ; \\
B_{j}(\omega)  & = & \frac{e\rho\epsilon^{3ij}}{\Rb^{2}}\left( -i\omega
                      R^{i}_{0}(\omega) \right) .
\label{bjw}
\eea
Carrying out the Gaussian integral gives,
\be
F(\rho) = \frac{Y(\rho)}{X(\rho)} ,
\label{genf5}
\ee
where,
\bea
Y(\rho) & = & \exp\left[ c_{1}\rho + \frac{c_{2}\rho^{2}}{2}\int_{-\wb}^{\wb}
               \frac{d\omega}{2\pi}\omega^{2}\left\{ 
                \frac{f_{1}-\rho f_{2}}{f_{3} -\rho^{2}f_{4}} \right\}\right] , 
                  \label{yrho} \\
X(\rho) & = & \sqrt{\prod_{\omega} c_{3}(f_{3}-\rho^{2}f_{4})} ,
                   \label{xrho}
\eea
and,
\be
\begin{array}{lll}
c_{1}=\delta E_{-}[\bR_{0}] & \hspace{0.5in} ; \hspace{0.5in} &  
                           f_{1}=|\bR_{0}(\omega)|^{2}(M^{2}\omega^{4}
                                     +\eta^{2}\omega^{2}) \\
c_{2}=\eta kT\hbar^{2}/{\cal T}^{2}\Rb^{4} & \hspace{0.5in} ; \hspace{0.5in} & 
                                   f_{2} = i\omega\, [ \hat{\bz}\cdot
                                      \bR_{0}(\omega)\times\bR^{\ast}_{0}
                                       (\omega)] (\eta kT\hbar /\ct\Rb^{2}) \\
c_{3}=M\Rb^{2}/2\eta^{2}kT & \hspace{0.5in} ; \hspace{0.5in} & 
                      f_{3}=(M^{2}\omega^{4}+\eta^{2}\omega^{2}
                                        )^{2} \\
         & {} \hspace{0.5in} {} \hspace{0.5in}  & 
                        f_{4}=(\eta kT\hbar\omega /\ct\Rb^{2})^{2} .
\label{consts}
\end{array}
\ee

\subsection{Average Energy-Level Shift}
\label{sec6c}
From eqn.~(\ref{ebar}),
\bea
\overline{\delta E_{-}} & = & \frac{d}{d\rho}[\ln Y(\rho)]\left. 
                               \right|_{\rho =0} - \frac{d}{d\rho}
                                [\ln X(\rho)]\left. \right|_{\rho =0}
            \nonumber \\
 & = & c_{1} - 0 \nonumber \\
 & = & \delta E_{-}[\bR_{0}] .
\label{eshft}
\eea
In the low-noise limit we find that the average energy-level shift is
given by the shift produced by the noiseless component of $\bR (t)$. The
absence of a further noise correction to eqn.~(\ref{eshft}) is surely a 
consequence of the low-noise approximation. Such a noise correction is 
expected to occur for stronger noise.

\subsection{Average Energy-Level Broadening}
\label{sec6d}
From eqn.~(\ref{var}),
\bea
\sigma^{2} & = & \frac{d^{2}}{d\rho^{2}}[\ln Y(\rho)]\left.\right|_{\rho =0}
                   - \frac{d^{2}}{d\rho^{2}}[\ln X(\rho)]\left.
                    \right|_{\rho =0} \nonumber \\
 & = & c_{2}\int_{\wc}^{\wb}\frac{d\omega}{\pi} \frac{\omega^{2}f_{1}}{f_{3}}
         + \ct\int_{\wc}^{\wb}\frac{d\omega}{\pi}\frac{f_{4}}{f_{3}} .
\label{fvar}
\eea
Solving the homogeneous Langevin equation for $\bR_{0}(t)$, and Fourier
transforming to obtain $\bR_{0}(\omega)$; introducing a dimensionless
frequency $x=\omega (M/\eta )$; and using eqn.~(\ref{consts}) gives
(eventually),
\be
\sigma = \frac{\hbar}{{\cal T}}
             \sqrt{ \left(\frac{MkT}{\eta^{2}\Rb^{2}}\right)\left[
              \frac{|\bR_{f}-\bR_{i}|^{2}}{\Rb^{2}}+
               \frac{kT}{(\eta\Rb^{2}/{\cal T})}\right]
                \int_{x_{c}}^{x_{b}}\frac{dx}{\pi}
                 \frac{1}{x^{2}(1+x^{2})^{2}} } .
\label{fvar2}
\ee

A simpler relation can be found if we choose boundary conditions such that,
\begin{displaymath}
|\bR_{f}-\bR_{i}|^{2} \sim \frac{2kT}{\eta}\ct ,
\end{displaymath}
which corresponds to fixing the separation of boundary points to be roughly the
same size as the diffusion cloud for Brownian motion over a time $\ct$. 
Eqn.~(\ref{fvar2}) becomes,
\be
\sigma \sim kT\left(\frac{\hbar}{\eta\Rb^{2}}\right) \kappa ,
\label{broad}
\ee
where,
\begin{displaymath}
\kappa = \sqrt{\frac{3\tau}{\ct}\int_{x_{c}}^{x_{b}}\frac{dx}{\pi}
          \frac{1}{x^{2}(1+x^{2})^{2}}} ,
\end{displaymath}
and $\tau = M/ \eta$. 
As mentioned above, eqns.~(\ref{fvar2}) and (\ref{broad}) should be thought of 
more as rough estimates, than as rigorous results.
Clearly though, we do see that level-broadening is
produced by the combination of Berry phase and stochastic effects. 
A Monte Carlo evaluation of the
path integrals appearing in eqns.~(\ref{ebar}) and (\ref{var}) would be
very interesting since such a calculation would not require the 
simplifications we found it necessary to make to produce an analytically
 tractable calculation. It is important to keep in mind that such a 
Monte Carlo calculation must still satisfy the third condition imposed in 
Section~\ref{sec3} to insure that the noise respects the adiabatic 
requirements of Berry's phase. 

\section{Closing Remarks}
\label{sec7}
In the usual Berry phase scenario one considers a pair of interacting
systems with vastly different dynamical time scales, and treats the
coupled dynamics using the Born-Oppenheimer approximation. In this
paper we generalize this scenario, allowing the slow system dynamics
to be stochastic as well as adiabatic. We introduce a model that allows us 
to study how the usual Berry phase scenario is modified by the stochastic
dynamics. 

Our principal results are: (1) a broadening and shifting of the fast system
energy-levels by a combination of Berry phase and stochastic effects; and
(2) the semiclassical limit of the slow system effective dynamics obeys
a Langevin equation in which geometrical effects produce Lorentz-like
and electric-like forces.

In the semiclassical and low-noise limit, we calculate approximately the
average level-shift and broadening produced by this geometric mechanism.
Monte Carlo evaluation of eqns.~(\ref{ebar}) and (\ref{var}) would be very 
interesting as this would free the analysis from the low-noise limit.

Formally, the semiclassical limit was taken by letting $\hbar \rightarrow 0$.
In fact, $\hbar$ is finite, and an experimental realization of this limit
must be approached differently. Formally, the semiclassical limit is 
controlled by the real exponential factor appearing in eqn.~(\ref{dyno}),
and is approached when it becomes sharply peaked. This occurs when
$\hbar\rightarrow 0$, though more generally when $\eta\Rb^{2}kT\ct\gg
\hbar^{2}$. Thus strong damping is one way to produce semiclassical behavior.
The low-noise limit required $kT \ll \eta\Rb^{2}/\ct$. Thus, sufficiently
low temperature will produce low-noise behavior. Our approximate results
for the average level-shift and broadening are thus expected to apply in
the limit of strong damping and sufficiently low temperature.

In an interesting Comment, Simon and Kumar \cite{sik} propose a
physical setting in which Berry's phase should produce level-broadening.
Their underlying idea is similar to the one we propose: a range of
level-shifts occur producing a broadening of the original energy level.
They do not provide a formal development of their proposal, however.
Gamliel and Reed \cite{gam} consider the original Berry phase scenario
of a pseudo-spin (fast system) interacting with an adiabatically
evolving pseudo-magnetic field (slow system). However, they allow for the
presence of a stochastic process whose sole effect is to relax the
pseudo-spin to an equilibrium state (which might evolve with time).
The stochastic process is assumed to produce {\em no\/} Berry phase
effects in the pseudo-spin dynamics. The deterministic motion of the
pseudo-magnetic field produces a unique level-shift $\delta E[\bR ]$
in each fast system energy level $E$. The central question for these
authors is whether the shift $\delta E[\bR ]$ is observable in the
presence of the conventional level-broadening produced by the relaxation
process. In our scenario, the slow system motion is the random
process, and the level-broadening arises through the Berry phase induced
level-shifts.

\section*{Acknowledgments}
It is a pleasure to thank the T-11 group at Los Alamos National Laboratory for 
the hospitality and support they provided during the time in which this work
was done. I would also like to thank T. Howell III for continued
support.

\appendix
\section*{}
\label{appendixa}
In this Appendix we consider a free classical Brownian particle moving in
one-spatial dimension. We will establish a condition that insures that the
Brownian motion is adiabatic relative to a quantum system that interacts
with the Brownian particle.

The Brownian motion is described by Langevin's equation,
\begin{equation}
m\ddot{R} + \eta\dot{R} = F(t) \vspace{0.1in} ,
\label{lan}
\end{equation}
where $m$ is the particle mass; $\eta$ is the friction coefficient; and
$F(t)$ is a stochastic force with correlation function,
\begin{equation}
\langle\: F(t)F(t^{\prime})\:\rangle = 2\eta kT\delta (t-t^{\prime})
  \hspace{0.1in} .
\label{cor}
\end{equation}
The stochastic force $F(t)$ fluctuates rapidly. The duration $\tau_{c}$
of a force fluctuation is known as the correlation time. It is determined
by the time scale of the microscopic processes which produce
the stochastic force. The particle velocity $\dot{R}$ varies on a much
slower time scale due to the large inertia of the Brownian particle.

Since the Brownian particle is free, $E=mv^{2}/2$. The time-averaged energy
is
\begin{displaymath}
\langle\: E \: \rangle = \frac{m}{2}\langle\: v^{2}\: \rangle =
    \frac{m}{2}K_{v}(0) ,
\end{displaymath}
where $K_{v}(t)=\langle\: v(0) v(t)\: \rangle$ is the velocity correlation
function. Its Fourier transform is the spectral density $J_{v}(\omega)$
of $v(t)$ \cite{wan},
\begin{displaymath}
K_{v}(t) = \int_{0}^{\infty}\, d\omega\, \cos\omega t\, J_{v}(\omega)
  \hspace{0.1in} .
\end{displaymath}
Thus,
\begin{equation}
\langle\, E \, \rangle = \frac{m}{2} \int_{0}^{\infty}\, d\omega
 \, J_{v}(\omega) \hspace{0.1in} . \label{eng}
\end{equation}
Defining the energy spectral density $\rho_{E} (\omega)$ as the mean 
noise-energy in the frequency range ($\omega$, $\omega + d\omega$),
we see from eqn.~(\ref{eng}) that, 
\begin{equation}
\label{den}
\rho_{E}(\omega) = \frac{m}{2} J_{v}(\omega) \hspace{0.1in} .
\end{equation}
It is well-known \cite{wan} that, $J_{v}(\omega) = \lim_{{\cal T}\rightarrow
\infty} 2|v(\omega)|^{2}/{\cal T}$. Since $v(\omega) = -i\omega R(\omega)$,
eqn.~(\ref{den}) becomes,
\begin{equation}
\rho_{E}(\omega) = \frac{m\omega^{2}}{2} J_{R}(\omega) \hspace{0.1in} ,
\label{nden}
\end{equation}
where $J_{R}(\omega) = \lim_{{\cal T}\rightarrow\infty} 2|R(\omega)|^{2}
/{\cal T}$.

It is preferable to express $\rho_{E}(\omega)$ in terms of $J_{F}(\omega)$.
To do this we use the Langevin equation to relate $J_{R}(\omega)$ to
$J_{F}(\omega)$. Fourier transforming eqn.~(\ref{lan}) gives, 
\begin{displaymath}
R(\omega) = \frac{-F(\omega)}{m\omega\left(\omega + i\omega_{r}\right)}
 \hspace{0.1in} ,
\end{displaymath}
where $\omega_{r}= \eta/m$. Thus,
\begin{displaymath}
J_{R}(\omega)= \frac{J_{F}(\omega)}{m^{2}\omega^{2}(\omega^{2}+
                     \omega_{r}^{2})} \hspace{0.1in} ,
\end{displaymath}
and,
\begin{displaymath}
\rho_{E}(\omega) = \frac{J_{F}(\omega)}{2m\left( \omega^{2} + \omega_{r}^{2}
 \right)} \hspace{0.1in} .
\end{displaymath}
In writing the force correlation function in eqn.~(\ref{cor}) as 
being proportional
to $\delta(t-t^{\prime})$, we assumed $\tau_{c}$ was effectively zero, and
$J_{F}(\omega)= 2\eta kT$ for all $\omega$. In fact, $\tau_{c}$ is not zero, 
though macroscopically small. Thus $J_{F}(\omega) = 2\eta kT$ only up to a 
cutoff frequency $\Lambda\sim 1/\tau_{c}$ so that
\begin{equation}
\label{den2} 
\rho_{E}(\omega) = \frac{\eta kT}{m\left( \omega^{2}+\omega_{r}^{2}\right)}
 \Theta (\Lambda - \omega ) \hspace{0.1in} ,
\end{equation}
where $\Theta(x)$ vanishes if $x< 0$ and is $1$ otherwise. From 
eqns.~(\ref{eng})---(\ref{nden}), and (\ref{den2}) we see that 
noise fluctuations in the random process $R(t)$ only exist for frequencies
$\omega\leq 1/\tau_{c}$. If the Brownian particle is coupled to a quantum 
system whose energy-level spacing is $\Delta E$, the Brownian motion will
be unable to produce transitions in the quantum system if $\hbar /\tau_{c}
\ll \Delta E$. If this condition is satisfied, the quantum system will see
the Brownian motion as adiabatic. This is the desired adiabaticity condition.

\end{document}